\pdfoutput=1
\documentclass[10pt,twocolumn]{article}

\usepackage[margin=0.72in]{geometry}
\usepackage{amsmath,amssymb}
\usepackage{booktabs}
\usepackage{graphicx}
\usepackage{microtype}
\usepackage[raggedright]{titlesec}
\usepackage[numbers,sort&compress]{natbib}
\usepackage{placeins}
\usepackage{enumitem}
\usepackage[table]{xcolor}
\usepackage{tikz}
\usepackage{url}
\usepackage[hidelinks]{hyperref}
\usetikzlibrary{arrows.meta,positioning}

\newcommand{\QCA}{QCA}
\newcommand{\vect}[1]{\mathbf{#1}}
\title{Compact Quantum Channel Attention for Static ASL Recognition:\\
A Controlled Study with Parameter-Matched Baselines}
\author{\normalsize
\begin{tabular}[t]{c}
Ronak Bhalgami\\
CSPIT, Charotar University\\
of Science and Technology\\
\href{mailto:ronak42.b@gmail.com}{ronak42.b@gmail.com}
\end{tabular}\hspace{1.5em}%
\begin{tabular}[t]{c}
Dr.~Deep Kothadiya\\
SCIT, Symbiosis\\
International University\\
\href{mailto:deep@scit.edu}{deep@scit.edu}
\end{tabular}\hspace{1.5em}%
\begin{tabular}[t]{c}
Dr.~Nikita Bhatt\\
CSPIT, Charotar University\\
of Science and Technology\\
\href{mailto:nikitabhatt.ce@charusat.ac.in}{nikitabhatt.ce@charusat.ac.in}
\end{tabular}%
}
\date{}

\begin{document}
\maketitle

\begin{abstract}
Hybrid quantum-classical models often report good accuracy on image tasks,
but it is rarely clear whether the quantum part is responsible. This paper
proposes QCA-DenseNet, a static American Sign Language (ASL) recognition
pipeline in which an eight-qubit variational circuit acts as a
channel-attention gate on DenseNet121 features. The quantum channel attention
(\QCA{}) gate has 17,464 parameters, 7.51 times fewer than a
squeeze-and-excitation (SE) block. The pipeline is tested on 1,815 MU
HandImages images by holding out one of five signers at a time, and a
frozen-backbone ablation uses small parts of the Kaggle ASL Alphabet data. On
unseen signers, QCA-DenseNet reaches 94.50\% mean accuracy, above SE
(93.72\%) and within 0.28 points of the best model. A matched classical MLP
and a model without attention perform about the same, and no difference is
significant. \QCA{} gives a much smaller attention gate without a detectable
loss in accuracy, but no quantum advantage.
\end{abstract}


\section{Introduction}
\label{sec:intro}

Fingerspelling is used to spell names and other words letter by letter, and
the recognition of static handshapes is a common starting point for
automatic sign language recognition \citep{rastgoo_2021_survey,wadhawan_2020_static}.
On the popular Kaggle ASL Alphabet dataset, convolutional networks already
reach more than 99.9\% test accuracy \citep{alsharif_2023_asl}. Such
results come from random image splits, where nearly identical frames of the
same person and background appear in both the training and the test set. A
harder and more useful test is a person that the model has never seen,
because hand size, skin tone, and camera setup change from one signer to the
next \citep{erdogan_2024_signer,sahoo_2022_mu}.

Variational quantum algorithms train a parameterised circuit with a
classical optimiser, and they are the main way to use present-day noisy
quantum devices for learning \citep{cerezo_2021_vqa,bharti_2022_nisq}. In
hybrid transfer learning, a pretrained convolutional network first
compresses the image and a small circuit then processes the features
\citep{mari_2020_transfer}. Several hybrid vision models of this kind report
good accuracy \citep{liu_2021_hqcnn,long_2025_hqcqcnn}. The difficulty is
attribution. A high score may come from the pretrained features, from the
classical layers around the circuit, or from a larger number of parameters.
Raubitzek and Mallinger found that quantum classifiers tuned by
hyperparameter search still trail modern boosting methods
\citep{raubitzek_2023_applicability}, and Schuld and Killoran argue that
advantage should not be the field's only goal \citep{schuld_2022_advantage}.

Channel attention is a convenient place to test a circuit under fair
conditions. A squeeze-and-excitation (SE) block reweights feature channels
through a small bottleneck network \citep{hu_2020_se}, and later designs
show that such reweighting can be done with far fewer parameters
\citep{wang_2020_eca,guo_2022_attention}. The size of an SE block grows with
the number of channels, while the trainable weights of a variational circuit
grow only with the number of qubits and layers. A gate built around a small
circuit is therefore much smaller than SE. Two questions follow. Does such a
compact gate hold on to the accuracy of SE when the test signer is new, and
if it does, is the credit due to the circuit or to the classical layers
wrapped around it?

This paper answers both questions with one proposed pipeline, QCA-DenseNet,
and three reference modules that remove one part at a time: an SE block with
more capacity, a classical MLP with the same size as \QCA{}, and a model
with no attention. The main test trains every DenseNet121 layer and holds
out one signer at a time. A frozen-backbone test with little training data
then looks at the attention head alone. The contributions are listed below.
\vspace{0pt plus 1fil}
\begin{itemize}[itemsep=8pt plus 1fil, topsep=4pt, parsep=0pt, partopsep=0pt]
  \item C1: An eight-qubit, two-layer variational channel gate that needs
  17,464 attention parameters, 7.51 times fewer than SE
  (Section~\ref{sec:method}, checked in Section~\ref{sec:res-params}).
  \item C2: Under leave-one-signer-out testing with full fine-tuning,
  QCA-DenseNet stays within 0.28 points of the best model and scores above
  SE, with no significant paired difference (Section~\ref{sec:res-mu}).
  \item C3: A matched MLP, a model without attention, signer-level tests,
  and a three-seed frozen test show that neither the circuit nor the
  bottleneck adds measurable accuracy. The accuracy comes from the
  fine-tuned backbone (Sections~\ref{sec:res-attr} and~\ref{sec:res-frozen}).
\end{itemize}

\section{Related Work}
\label{sec:related}

\subsection{Static Sign Recognition}

Deep convolutional networks recognise isolated static signs with high
accuracy when the training and test images come from the same collection
\citep{wadhawan_2020_static}. Alsharif et al.\ compared AlexNet, ConvNeXt,
EfficientNet, ResNet-50, and a vision transformer on the 87,000-image Kaggle
ASL Alphabet set and reported 99.98\% accuracy for ResNet-50
\citep{alsharif_2023_asl}. At that level there is little room left to rank
new modules. A survey by Rastgoo et al.\ covers handshape, motion, and
multimodal methods \citep{rastgoo_2021_survey}, and Camgoz et al.\ used
transformers to recognise and translate continuous signing from video
\citep{camgoz_2020_slt}. Those video tasks contain timing and language
structure that single alphabet frames do not have. Erdo\u{g}an and
Bayta\c{s} showed that the identity of the signer is a separate source of
error \citep{erdogan_2024_signer}, and Sahoo et al.\ tested static hand
gestures on the MU dataset with leave-one-subject-out validation
\citep{sahoo_2022_mu}. The present study follows the same practice.

\subsection{Hybrid Quantum-Classical Vision Models}

Variational circuits are trained by classical optimisers that use measured
expectation values \citep{cerezo_2021_vqa}. Mari et al.\ joined a pretrained
classical network with a final circuit for image classification
\citep{mari_2020_transfer}. Quanvolutional networks apply small random
circuits to image patches \citep{henderson_2020_quanvolutional}, and later
designs place trainable circuits inside or after classical feature
extractors \citep{liu_2021_hqcnn,hur_2022_qcnn,long_2025_hqcqcnn}. The
design with the closest name is the channel attention of Budiutama et al.,
but there the attention acts on quantum states inside a quantum
convolutional network that recognises quantum phases
\citep{budiutama_2024_channel}. In \QCA{} the input is classical throughout:
the circuit receives eight angles computed from the pooled channels of an
ordinary image backbone, and its output only rescales those channels.

Theory gives reasons to expect little from a circuit this small. Schuld et
al.\ showed that such models are partial Fourier series whose frequencies
are fixed by the encoding gates, so a single $R_Y$ encoding per input yields
only a first-degree series in that input \citep{schuld_2021_encoding};
repeating the encoding between layers, as in data re-uploading, widens the
spectrum \citep{schuld_2021_encoding,perez_2020_reupload}. Capacity and
generalisation bounds tie what these models can learn to the number of
trainable gates \citep{abbas_2021_power,caro_2022_generalization}, and
Huang et al.\ found that classical learners given access to the same data
can often match quantum models \citep{huang_2021_data}. Trainability is a
separate problem. Gradients can vanish as circuits grow or when the cost
function is global \citep{cerezo_2021_barren}, and hardware noise can
flatten them in the same way \citep{wang_2021_noise}.

\subsection{Channel Attention and Fair Comparison}

Hu et al.\ built SE as a two-layer bottleneck that squeezes each channel to
its global average and then rescales the channels with sigmoid weights
\citep{hu_2020_se}. ECA-Net replaces the bottleneck with a one-dimensional
convolution across neighbouring channels and reports accuracy similar to or
better than SE with only a handful of parameters \citep{wang_2020_eca}. Guo
et al.\ survey these and many other attention mechanisms used in vision
\citep{guo_2022_attention}. On the quantum side, Raubitzek and Mallinger
ask that settings be stated clearly and that classical baselines be
comparable \citep{raubitzek_2023_applicability}, and Schuld and Killoran
suggest that the field would learn more by asking where circuits help than
by searching for an advantage \citep{schuld_2022_advantage}. A broader
warning comes from Bouthillier et al., who showed that the data sample and
the random seed alone can reorder methods in a benchmark
\citep{bouthillier_2021_variance}. The matched MLP in this study answers the
call for comparable baselines, and the repeated seeds and per-signer tests
answer the variance problem.

\section{Proposed QCA-DenseNet}
\label{sec:method}

\subsection{Pipeline Overview}

\begin{figure*}[!t]
\centering
\resizebox{0.97\textwidth}{!}{\begin{tikzpicture}[
  font=\footnotesize,
  box/.style={draw, rounded corners=2pt, align=center, minimum height=11mm,
              inner xsep=4pt},
  cls/.style={box, fill=blue!8, draw=blue!55!black},
  qnt/.style={box, fill=violet!12, draw=violet!65!black},
  gat/.style={box, fill=orange!14, draw=orange!70!black},
  flow/.style={-{Latex[length=2mm]}, thick, draw=gray!70!black},
  wire/.style={draw=gray!60!black, line width=0.5pt},
  sel/.style={draw=violet!65!black, fill=violet!18, rounded corners=1pt},
  ry/.style={draw=violet!65!black, fill=white, inner sep=1pt,
             minimum width=8mm, minimum height=2.6mm, font=\tiny},
  meas/.style={draw=violet!65!black, fill=white, inner sep=1pt,
               minimum width=6mm, minimum height=2.6mm, font=\tiny}
]
\node[cls] (img) at (0,0) {Hand image\\$224\times224$};
\node[cls, right=6mm of img] (cnn) {DenseNet121\\(fine-tuned)};
\node[cls, right=6mm of cnn] (gap) {Global pooling\\$\mathbf{x}\in\mathbb{R}^{1024}$};
\node[qnt, right=6mm of gap, minimum width=26mm] (qca) {\textbf{QCA gate}\\$\mathbf{a}\in(0,1)^{1024}$};
\node[gat, right=6mm of qca] (mul) {Reweighting\\$\mathbf{x}\odot\mathbf{a}$};
\node[cls, right=6mm of mul] (fc) {Linear\\classifier};
\node[cls, right=6mm of fc] (out) {Class label\\(26 or 29)};
\draw[flow] (img) -- (cnn);
\draw[flow] (cnn) -- (gap);
\draw[flow] (gap) -- (qca);
\draw[flow] (qca) -- (mul);
\draw[flow] (mul) -- (fc);
\draw[flow] (fc) -- (out);
\draw[flow] (gap.north) to[out=35,in=145]
  node[above, font=\scriptsize, text=gray!70!black]{$\mathbf{x}$ passed to reweighting} (mul.north);
\node[anchor=west, font=\small\bfseries] at ([xshift=-2mm,yshift=11mm]img.north west) {(a)};

\begin{scope}[yshift=-38mm]
\node[cls] (xin) at (0,0) {$\mathbf{x}$\\1024 channels};
\node[cls, right=6mm of xin] (down) {$W_d$: $1024\!\rightarrow\!8$\\$\tanh(\cdot)\,\pi/2$};
\def\cx{6.0}
\foreach \i in {0,...,7} {
  \pgfmathsetmacro{\yy}{1.40-0.40*\i}
  \pgfmathtruncatemacro{\k}{\i+1}
  \draw[wire] (\cx,\yy) -- (\cx+6.4,\yy);
  \node[ry] at (\cx+0.75,\yy) {$R_Y(z_{\k})$};
  \node[meas] at (\cx+5.75,\yy) {$\langle Z\rangle$};
}
\draw[sel] (\cx+1.55,-1.58) rectangle (\cx+3.05,1.58);
\node[align=center, font=\scriptsize] at (\cx+2.30,0) {Strongly\\entangling\\layer 1};
\draw[sel] (\cx+3.40,-1.58) rectangle (\cx+4.90,1.58);
\node[align=center, font=\scriptsize] at (\cx+4.15,0) {Strongly\\entangling\\layer 2};
\draw[draw=violet!65!black, dashed, rounded corners=3pt]
  (\cx-0.15,-1.80) rectangle (\cx+6.55,1.80);
\node[font=\scriptsize, text=violet!65!black, anchor=south]
  at (\cx+3.2,1.80) {8-qubit variational circuit (simulated)};
\node[cls, anchor=west] (up) at (\cx+7.2,0) {$W_u$: $8\!\rightarrow\!1024$\\sigmoid};
\node[gat, right=6mm of up] (aout) {$\mathbf{a}$\\channel gates};
\draw[flow] (xin) -- (down);
\draw[flow] (down.east) -- (\cx-0.15,0);
\draw[flow] (\cx+6.55,0) -- (up.west);
\draw[flow] (up) -- (aout);
\node[anchor=west, font=\small\bfseries] at ([xshift=-2mm,yshift=16mm]xin.north west) {(b)};
\end{scope}
\end{tikzpicture}}
\caption{Proposed QCA-DenseNet pipeline (a) and the internal structure of the QCA gate (b).}
\label{fig:pipeline}
\end{figure*}
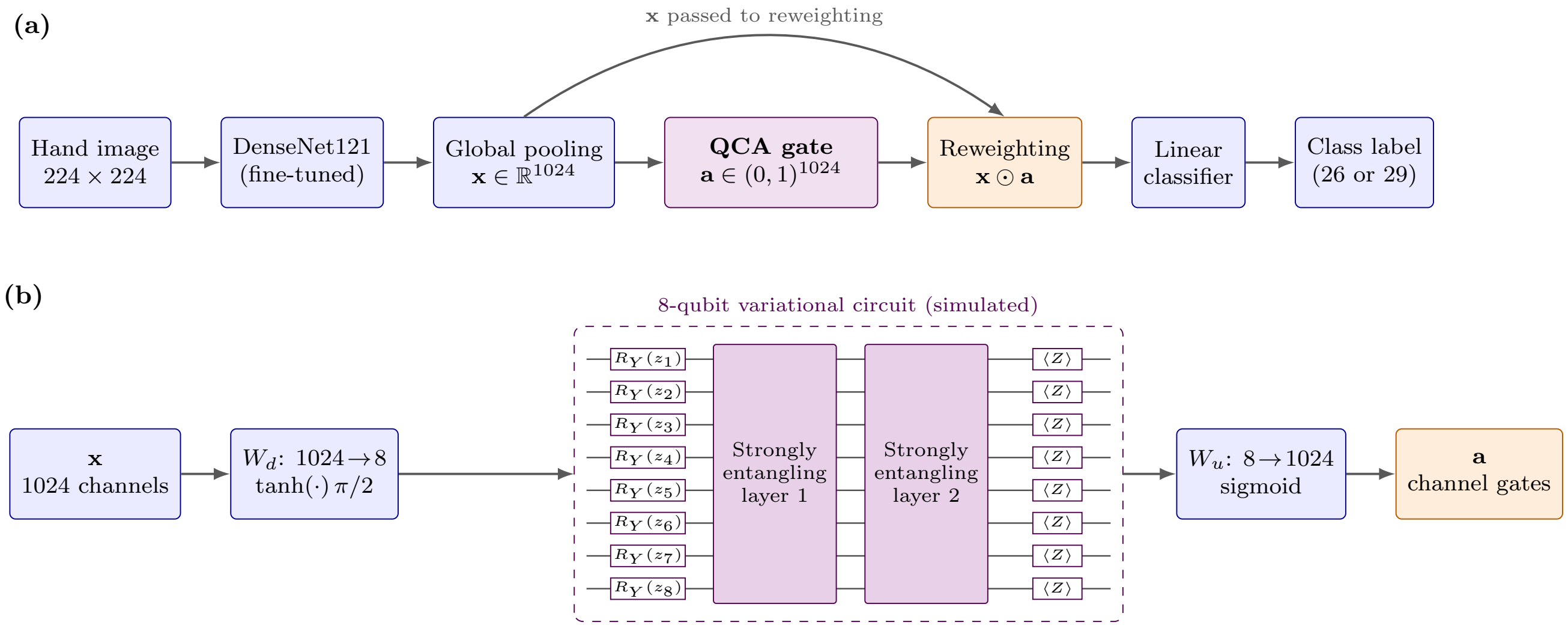

Figure~\ref{fig:pipeline}(a) follows one image through the pipeline.
DenseNet121 \citep{huang_2022_densenet}, initialised with ImageNet weights,
turns a $224\times224$ RGB hand image into feature maps, and global average
pooling reduces these to a vector $\vect{x}\in\mathbb{R}^{1024}$. The \QCA{}
gate reads $\vect{x}$ and returns one weight per channel,
$\vect{a}\in(0,1)^{1024}$, and the linear classifier sees only the product
$\vect{x}'=\vect{x}\odot\vect{a}$. In the main protocol the backbone, gate
and classifier are trained together with a cross-entropy loss; the ablation
in Section~\ref{sec:ablation} freezes the backbone. Since every weight lies
between 0 and 1, the gate can damp a channel but cannot amplify it or create
a new feature. Inside the gate (Figure~\ref{fig:pipeline}(b)), a classical
layer first compresses $\vect{x}$ to eight rotation angles. An eight-qubit
circuit with two entangling layers acts on these angles, and a second
classical layer expands the eight measured values back into 1,024 sigmoid
gates.

\subsection{Quantum Channel Attention Gate}

The reduced features are first bounded,
\begin{equation}
  \vect{z} = \frac{\pi}{2}\tanh\!\left(W_d\vect{x}+\vect{b}_d\right),
  \qquad W_d\in\mathbb{R}^{8\times1024},
  \label{eq:encode}
\end{equation}
so that every angle lies in $(-\pi/2,\pi/2)$. Each angle sets one $R_Y$
rotation, which gives the state
\begin{equation}
  |\psi(\vect{z},\boldsymbol{\theta})\rangle =
  U_2(\boldsymbol{\theta}_2)\,U_1(\boldsymbol{\theta}_1)
  \bigotimes_{i=1}^{8} R_Y(z_i)\,|0\rangle^{\otimes 8}.
  \label{eq:state}
\end{equation}
Each $U_\ell$ is a strongly entangling layer. It applies a general
three-angle rotation to every qubit and then a cyclic pattern of CNOT gates,
whose range grows with the layer number. Following Schuld et al., this
encoding fixes the frequencies that the circuit can express
\citep{schuld_2021_encoding}. The gate output is
\begin{equation}
  q_i = \langle\psi|Z_i|\psi\rangle,\qquad
  \vect{a} = \sigma\!\left(W_u\vect{q}+\vect{b}_u\right),
  \label{eq:gate}
\end{equation}
with $W_u\in\mathbb{R}^{1024\times8}$. The down layer has 8,200 parameters,
the circuit has 48 angles ($2\times8\times3$), and the up layer has 9,216,
for a total of 17,464. The circuit is simulated exactly with the PennyLane
\texttt{default.qubit} device and trained by backpropagation. The simulation
runs on the CPU, and the eight measured values are then moved back to the
accelerator.

\subsection{Reference Modules}

Three reference modules take the place of the \QCA{} gate, while the
backbone, classifier and training schedule stay the same. The \emph{matched
MLP} is \QCA{} with the circuit cut out: both projections remain, and a
bias-free $8\times8$ layer between two $\tanh$ functions stands in for the
circuit, so the two differ only in their eight-unit core. \emph{SE} follows
Hu et al.\ \citep{hu_2020_se} with bias-free layers, a reduction ratio of
16, a ReLU between the layers and sigmoid outputs. With \emph{no attention},
$\vect{x}$ goes directly to the classifier. Table~\ref{tab:parameters}
gives the parameter counts; the frozen column includes the 29-class Kaggle
classifier, and the full column refers to the 26-class MU model with every
layer trainable.

\begin{table}[!hb]
\centering
\caption{Parameter counts of all channel modules.}
\label{tab:parameters}
\resizebox{\columnwidth}{!}{%
\begin{tabular}{lrrr}
\toprule
Module & Attention & Trainable (frozen) & Trainable (full) \\
\midrule
\textbf{QCA (proposed)} & \textbf{17,464} & 47,189 & 6,997,970 \\
Matched MLP & 17,480 & 47,205 & 6,997,986 \\
SE ($r=16$) & 131,072 & 160,797 & 7,111,578 \\
No attention & 0 & 29,725 & 6,980,506 \\
\bottomrule
\end{tabular}}
\end{table}

\section{Experimental Setup}
\label{sec:setup}

\subsection{Datasets}

Kaggle's ASL Alphabet collection \citep{alsharif_2023_asl} holds 87,000
images in 29 classes, the 26 letters plus \emph{del}, \emph{nothing} and
\emph{space}. A stratified 80/10/10 split with seed 42
gives 8,700 validation and 8,700 test images. The separate 28-image test
folder is not used. The public mirror of the MU HandImages ASL dataset
\citep{sahoo_2022_mu} has 2,515 images of 36 signs from five people. Only the 1,815 images of the
letters A to Z are kept, which gives a 26-class task. The five signers
contribute 650, 645, 130, 130, and 260 images. Horizontal flips are not
used for augmentation, because a flip changes the hand used for the sign.

\begin{figure*}[!t]
\centering
\includegraphics[width=0.92\textwidth]{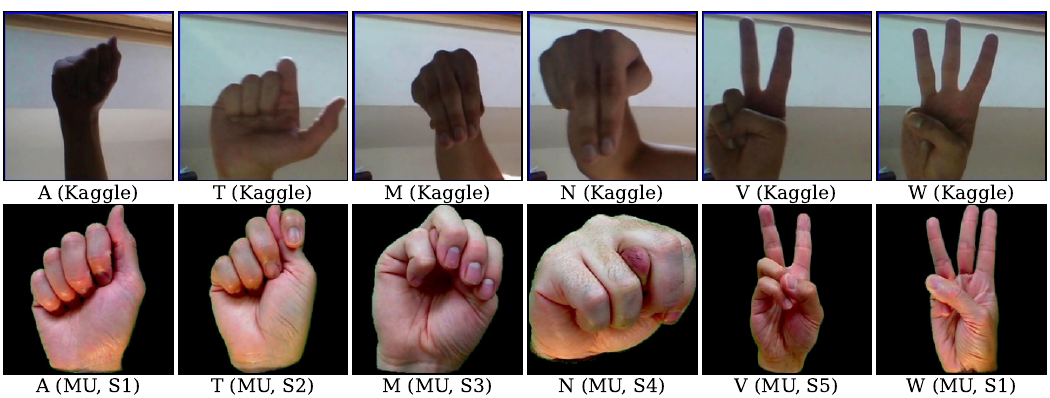}
\caption{Letters A, T, M, N, V, and W from Kaggle (top) and from different MU signers (bottom).}
\label{fig:samples}
\end{figure*}

Figure~\ref{fig:samples} shows six letters from both datasets. These are
the letters that the model confuses most often in Section~\ref{sec:res-mu}.
The Kaggle images share one indoor background and similar lighting, while
the MU images are cropped hands on a black background from different
people. A model that depends on the background or on one person's hand would
fail on the MU test, which is the concern raised by Erdo\u{g}an and
Bayta\c{s} \citep{erdogan_2024_signer}. In the pairs A/T and M/N only the
thumb position changes, and V and W differ by one raised finger.

\subsection{Main Protocol: Signer-Held-Out Testing}

Each of the five folds holds out one MU signer for testing. From the other
four signers, a
class-stratified 15\% of the images forms the validation set. No layer is
frozen. AdamW updates the backbone with a learning rate of $0.00001$ and the
attention module and classifier with $0.001$, using weight decay $0.0001$
and cosine annealing in both groups. Each fold runs for 15 epochs at batch
size 32 with seed 42, and the held-out signer is scored once, with the
checkpoint that reached the highest validation accuracy.

\subsection{Ablation: Frozen Backbone with Little Data}
\label{sec:ablation}

Fine-tuning the whole network on the random Kaggle split gives 100\% test accuracy
with and without \QCA{}, close to the ceiling that Alsharif et al.\ report
\citep{alsharif_2023_asl}, so this split cannot separate attention modules.
The ablation therefore keeps the ImageNet backbone frozen, leaving the head
as the only source of differences, and cuts the training portion
(stratified by class) to 1\%, 5\% or 10\% while validation and test keep
8,700 images each. Heads are trained for 15 epochs at batch size 32 with
AdamW, a learning rate of $0.001$ under cosine annealing and weight decay
$0.0001$; the whole grid is repeated with seeds 42, 43 and 44.

\subsection{Evaluation Metrics}

All metrics are computed again from the saved checkpoints. Accuracy and
macro-F1 are reported. For MU, the main summary is the mean and sample
standard deviation over the five signers, because pooled scores give more
weight to signers with more images, and the largest fold is five times the
smallest. Pooled accuracy is reported as a second metric. Two-sided paired
$t$-tests compare \QCA{} with each reference module over the five signer
accuracies, without correction for multiple tests. With only five pairs the
tests have low power, so a non-significant result does not prove that two
models are equal. Kaggle results are averaged over three seeds, because
random initialisation alone can change the order of close methods
\citep{bouthillier_2021_variance}.

\section{Results}
\label{sec:results}

\subsection{Parameter Efficiency of the QCA Gate (C1)}
\label{sec:res-params}

Table~\ref{tab:parameters} confirms the first contribution. The \QCA{} gate
uses 17,464 attention parameters and SE uses 131,072, so \QCA{} needs 7.51
times fewer, or 13.3\% of the SE budget. Only 48 of the 17,464 parameters
are circuit angles. The rest belong to the two projections, as in any
bottleneck gate \citep{hu_2020_se,wang_2020_eca}. The matched MLP differs from
\QCA{} by 16 parameters, so comparing the two isolates the circuit. With full
fine-tuning the backbone dominates the total count, so the attention column
is the one that matters.

\subsection{Accuracy on Unseen Signers (C2)}
\label{sec:res-mu}

Table~\ref{tab:mu} summarises the held-out-signer results. Averaged over the
five signers, QCA-DenseNet scores 94.50\%, which is 0.28 points behind the
model without attention (94.78\%, $p=0.543$) and 0.05 points behind the
matched MLP (94.55\%, $p=0.943$). Against SE the sign reverses: \QCA{} is
0.78 points ahead of SE (93.72\%) with 13.3\% of its attention parameters,
although this accuracy difference is not significant ($p=0.116$), and
macro-F1 moves the same way (94.26\% against 93.04\%). On pooled accuracy
\QCA{} ranks first, at 94.99\%. That ranking rests on the weighting: the two
large signers supply most of the pooled test images, and when each signer
counts equally the lead disappears. Table~\ref{tab:mu} thus supports parity
at a much lower attention cost; it does not show that \QCA{} is better. With
five signer pairs and a single training seed, even the parity is an
observation rather than a demonstrated equivalence
\citep{bouthillier_2021_variance}.

\begin{table}[!hb]
\centering
\caption{Held-out-signer MU results (\%).}
\label{tab:mu}
\resizebox{\columnwidth}{!}{%
\begin{tabular}{lcccc}
\toprule
Module & Signer acc. & Signer F1 & Pooled acc. & $p$ vs.\ QCA \\
\midrule
\textbf{QCA (proposed)} & $94.50 \pm 4.95$ & $94.26 \pm 4.94$ & \textbf{94.99} & ref. \\
Matched MLP & $94.55 \pm 4.02$ & $93.95 \pm 4.45$ & 94.88 & 0.943 \\
SE & $93.72 \pm 5.17$ & $93.04 \pm 5.43$ & 94.71 & 0.116 \\
No attention & $94.78 \pm 5.17$ & $94.61 \pm 5.28$ & 94.82 & 0.543 \\
\bottomrule
\end{tabular}}
\end{table}

Table~\ref{tab:signers} shows why the module differences are so hard to
separate. \QCA{} is ahead of SE on signers 3, 4 and 5, level on signer 2
and behind on signer 1. All four modules find signer 5 the hardest, with
accuracies between 84.62\% and 87.69\%, whereas the other folds stay between
94.77\% and 98.46\%. Changing the test person moves accuracy by
roughly ten points, while changing the attention module moves it by at most
3.08 points on any single signer. Erdo\u{g}an and Bayta\c{s} treat signer
identity as a separate source of error \citep{erdogan_2024_signer}; these
folds show how large that source is here.

\begin{table}[!hb]
\centering
\caption{Per-signer MU test accuracy (\%).}
\label{tab:signers}
\resizebox{\columnwidth}{!}{%
\begin{tabular}{lccccc}
\toprule
Module & S1 (650) & S2 (645) & S3 (130) & S4 (130) & S5 (260) \\
\midrule
\textbf{QCA (proposed)} & 95.38 & 97.52 & 96.92 & 96.92 & 85.77 \\
Matched MLP & 94.77 & 97.21 & 97.69 & 95.38 & 87.69 \\
SE & 95.69 & 97.52 & 95.38 & 95.38 & 84.62 \\
No attention & 95.38 & 96.59 & 98.46 & 97.69 & 85.77 \\
\bottomrule
\end{tabular}}
\end{table}

\begin{figure*}[!t]
\centering
\includegraphics[width=0.92\textwidth]{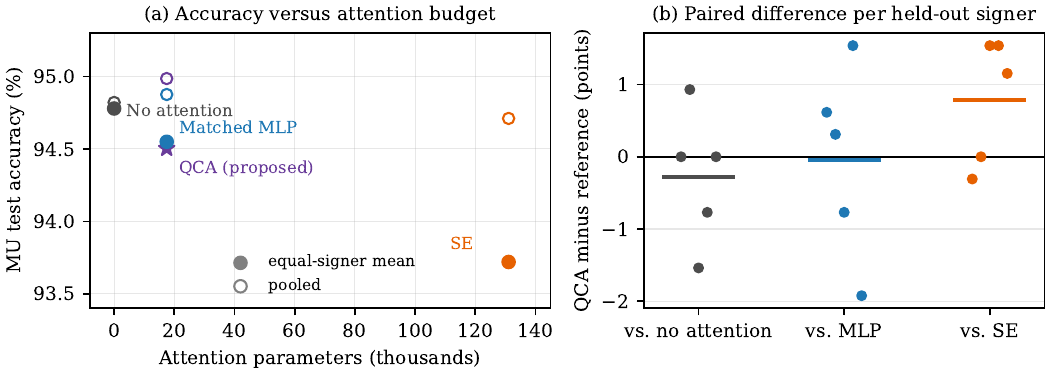}
\caption{MU accuracy versus attention parameters (a) and paired per-signer differences from QCA (b).}
\label{fig:efficiency}
\end{figure*}

Figure~\ref{fig:efficiency}(a) places each module by attention budget and
MU accuracy. The three small modules, none above 17,480 attention
parameters, sit within 0.28 points of one another, and SE, at 131,072
parameters, has the lowest signer mean. Figure~\ref{fig:efficiency}(b)
plots the five per-signer differences behind each $t$-test. Relative to no
attention and to the matched MLP they fall on both sides of zero; relative
to SE, four of the five are zero or positive. \QCA{} sits with the cheap
modules at no visible cost in accuracy, but it is not ahead of the
cheapest ones.

\begin{figure*}[!t]
\centering
\includegraphics[width=0.8\textwidth]{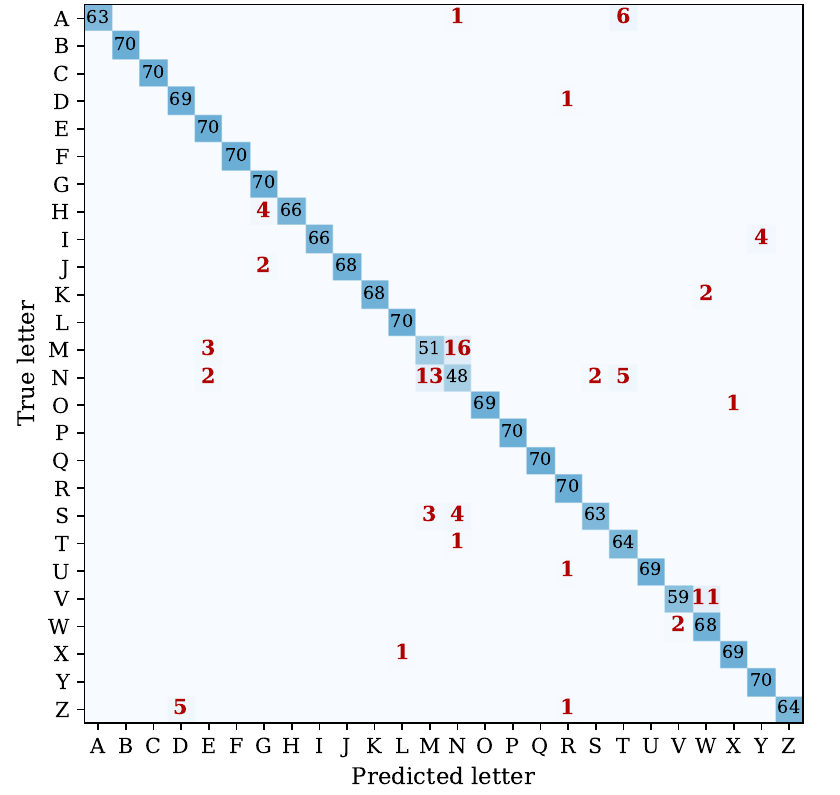}
\caption{QCA-DenseNet predictions pooled over the five held-out MU signers.}
\label{fig:cm}
\end{figure*}

Figure~\ref{fig:cm} pools the QCA-DenseNet predictions from all five
held-out signers, with correct predictions on the diagonal and errors in
red. Of the 1,815 test images, 1,724 are classified correctly, and most of
the mistakes involve a few pairs of similar handshapes. M and N are confused
in both directions (16 and 13 images), V is read as W 11 times, and A as T
6 times. Within each pair the letters differ only in the thumb
position or in how many fingers are raised or folded, a detail that is easy
to miss in one still frame.

\subsection{Contribution of the Circuit (C3)}
\label{sec:res-attr}

Because the matched MLP has every part of \QCA{} except the circuit,
comparing the two isolates what the circuit contributes. On MU they reach
94.55\% and 94.50\% (Table~\ref{tab:mu}), a gap of 0.05 points with
$p=0.943$, so replacing the circuit with an eight-unit classical layer makes
no difference that this test can detect. Dropping the gate altogether does
not hurt either: the model without attention, which has no gate
parameters, scores 94.78\% ($p=0.543$ against \QCA{}). Neither the circuit
nor the bottleneck around it adds measurable accuracy, and the performance
on unseen signers is best credited to the fine-tuned DenseNet121 features.
Huang et al.\ reached a related conclusion in a different setting, where
classical learners with access to data matched quantum models
\citep{huang_2021_data}.

\subsection{Frozen-Backbone Ablation (C3)}
\label{sec:res-frozen}

With the backbone frozen only the head can adapt, and
Table~\ref{tab:low-data} shows a different ordering. SE now leads at every
training fraction, in both accuracy and macro-F1, which fits its much larger
head. The three small heads change places from one fraction to the next. At
1\%, \QCA{} reaches 62.16\%, 1.36 points above the matched MLP, yet it
trails no attention (64.69\%) and SE (66.17\%). At 5\% the four means lie
within 1.09 points and \QCA{} is last. At 10\% it reaches 90.36\%, 0.17
points above no attention and 0.58 points below the matched MLP. Macro-F1 ranks the modules
exactly as accuracy does at each fraction. One extra frozen run on the full
training split (seed 42) keeps the order seen at 10\%: SE 97.10\%, \QCA{}
95.98\%, no attention 95.89\%.

\begin{table}[!hb]
\centering
\caption{Frozen Kaggle ablation (\%, mean $\pm$ SD).}
\label{tab:low-data}
\small
\begin{tabular}{llcc}
\toprule
Data & Module & Accuracy & Macro-F1 \\
\midrule
1\% & \textbf{QCA (proposed)} & $62.16 \pm 1.16$ & $60.49 \pm 1.46$ \\
 & Matched MLP & $60.80 \pm 1.31$ & $59.06 \pm 1.79$ \\
 & SE & $66.17 \pm 0.42$ & $65.25 \pm 0.11$ \\
 & No attention & $64.69 \pm 0.63$ & $63.42 \pm 0.92$ \\
\midrule
5\% & \textbf{QCA (proposed)} & $86.13 \pm 0.93$ & $86.06 \pm 0.94$ \\
 & Matched MLP & $86.39 \pm 0.47$ & $86.34 \pm 0.50$ \\
 & SE & $87.22 \pm 0.88$ & $87.18 \pm 0.90$ \\
 & No attention & $86.43 \pm 0.55$ & $86.38 \pm 0.56$ \\
\midrule
10\% & \textbf{QCA (proposed)} & $90.36 \pm 0.35$ & $90.35 \pm 0.38$ \\
 & Matched MLP & $90.93 \pm 0.24$ & $90.94 \pm 0.23$ \\
 & SE & $91.12 \pm 0.45$ & $91.13 \pm 0.46$ \\
 & No attention & $90.19 \pm 0.60$ & $90.17 \pm 0.62$ \\
\bottomrule
\end{tabular}
\end{table}

A data re-uploading variant \citep{perez_2020_reupload}, which repeats the
angle encoding before each variational layer, was also tried at 1\% data
with seed 42 and the same 17,464 parameters. It reached 61.38\% test
accuracy, slightly below the 61.83\% of the standard gate trained with that
seed. The wider frequency spectrum did not turn into accuracy in this
single-seed pilot, and the variant was not pursued further.

Measured against the contributions listed in Section~\ref{sec:intro}, C1
holds by construction, C2 holds as comparable rather than better accuracy,
and C3 finds no measurable effect of the circuit.

\section{Discussion}
\label{sec:discussion}

SE leads every frozen comparison and yet has the lowest signer mean on MU.
With a frozen backbone the head is the only trainable part, so extra head capacity pays
off, and SE has far more of it than the other three modules. Once every
layer is trained, the backbone can learn hand features on its own, and the
gate is probably left to make small corrections to features that already
suit the task. Under those conditions the 17,464-parameter gate did no worse
than the 131,072-parameter one, and neither moved far from the model with
no gate at all. The 48 circuit angles behaved, in effect, like a classical
bottleneck of the same width. Scaling the circuit up in the hope of a
quantum advantage would come at a price, since larger circuits tend to be
harder to train.

How the MU results are averaged also changes the ranking. Pooled accuracy
describes the particular mix of images in MU, whereas the signer mean
estimates what a new person can expect, and \QCA{} leads only on the first.
Reporting the pooled 94.99\% alone would imply a ranking that the
signer-level tests do not support. Parameter counts need similar care. The 113,608
parameters saved relative to SE amount to about 1.6\% of the full model,
and wall-clock time was not compared. A smaller gate is therefore mainly
of interest when the attention budget itself is tight, and the frozen
ablation shows that in that setting it gives up some accuracy.

\section{Future Scope}
\label{sec:future}

Future work will add more signers and training seeds, compare \QCA{} with
ECA-Net, test the circuit under noise or on quantum hardware, and extend the
pipeline to continuous fingerspelling.

\FloatBarrier
\section{Conclusion}
\label{sec:conclusion}

This paper proposed QCA-DenseNet, a static ASL recognition pipeline in which
an eight-qubit variational circuit gates the pooled channels of DenseNet121.
The \QCA{} gate needs 17,464 attention parameters, 7.51 times fewer than SE.
In leave-one-signer-out testing with full fine-tuning, the pipeline reached
94.50\% mean accuracy over signers, within 0.28 points of the best model and
above SE, and it had the highest pooled accuracy (94.99\%). A matched MLP
and a model without attention performed about the same, no paired difference
was significant, and SE was stronger whenever the backbone was frozen.
\QCA{} is therefore a compact gate that is comparable to, but does not
improve on, classical options. The accuracy of the pipeline comes from the fine-tuned
backbone, and no quantum advantage was observed.

\section*{Author Contributions}
\textbf{Ronak Bhalgami:} Conceptualization, Methodology, Software,
Investigation, Formal analysis, Visualization, Writing -- original draft.
\textbf{Deep Kothadiya:} Supervision, Methodology, Writing -- review \&
editing. \textbf{Nikita Bhatt:} Supervision, Writing -- review \& editing.

\bibliographystyle{unsrtnat}
\bibliography{references}

\end{document}